\documentclass[aps,prb,twocolumn,superscriptaddress,amsmath,floatfix]{revtex4}

\usepackage{graphicx}
\usepackage{dcolumn}
\usepackage{bm}
\usepackage{natbib}

\begin{document}

\title{\textbf{
Dynamically-Induced Frustration as a 
Route to a Quantum Spin Ice State in
Tb$_2$Ti$_2$O$_7$ via Virtual Crystal Field Excitations
and Quantum Many-Body Effects}}

\author{Hamid R. Molavian}
\affiliation{
Department of Physics and Astronomy, University of Waterloo, Ontario, Waterloo N2L 3G1, Canada}

\author{Michel J.P. Gingras}
\affiliation{
Department of Physics and Astronomy, University of Waterloo, Ontario, Waterloo N2L 3G1, Canada}
\affiliation{
Department of Physics and Astronomy, University of Canterbury,
Private Bag 4800, Christchurch, New Zealand}

\author{Benjamin Canals}
\affiliation{
Laboratoire Louis N\'eel, CNRS, BP 166, 38042 Grenoble, France}

\date{\today}

\begin{abstract} 

The Tb$_2$Ti$_2$O$_7$ pyrochlore magnetic material is attracting much
attention for its {\em spin liquid} state, failing to develop long
range order down to 50 mK despite a Curie-Weiss temperature
$\theta_{\rm CW} \sim -14$ K. In this paper 
we reinvestigate the theoretical description 
of this material by considering 
a quantum model of independent tetrahedra 
to describe its low temperature properties.
The naturally-tuned proximity of
this system near a N\'eel to spin ice phase
boundary allows for 
a resurgence of quantum fluctuation effects that lead to
an important renormalization of its effective 
low energy spin Hamiltonian.
As a result, Tb$_2$Ti$_2$O$_7$ 
is argued to be a
{\em quantum spin ice}.
We put forward an experimental
 test of this proposal
using neutron scattering on a single crystal.
\end{abstract}

\maketitle

Magnetic frustration arises when the lattice geometry prevents 
a system from finding its
classical ground state energy by minimizing
the energy between pairs of interacting magnetic moments (spins),
pair by pair.  Particularly interesting are 
models of geometrically frustrated
magnets where there exists
a macroscopic number of classical ground states 
not related by any global symmetry~\cite{Moessner}.
A prominent class of such systems are the {\it spin ices} where
Ising spins reside on a three-dimensional pyrochlore
lattice of
corner-sharing tetrahedra~\cite{Harris-PRL,Bramwell-JPC,Bramwell-Science}.
Because of their macroscopic number of
quasi-degenerate low-energy states, spin ice materials possess
an extensive low-temperature 
magnetic entropy~\cite{Ramirez,Hertog,Cornelius}
 similar to that found in the proton disorded
phase of common water 
ice~\cite{Pauling}.

A current
and exciting direction of research in frustrated magnetism is 
the study of
low energy effective Hamiltonians and gauge 
theories~\cite{Hermele,Bergman,Pujol}
to describe highly frustrated systems which,
when ignoring quantum effects, 
display an extensive classical ground state degeneracy
similarly to 
spin ices.
Despite the seemingly broad conceptual context of gauge
theory approaches, there have so far been few real frustrated
quantum magnetic materials identified 
as potential candidates for the exotic behaviors proposed
by these theories~\cite{Bergman}.
In this paper we argue that the paradoxical 
Tb$_2$Ti$_2$O$_7$ (TTO) pyrochlore 
\cite{Gardner-PRL,Gingras-PRB,Gardner-PRB,Gardner-PRB-50mK,TTO,TTO-neutron,Reviews}
belongs to such a class of materials.
Specifically, we use a simple model to illustrate that the starting
point of the above theories, namely a frustrated Ising spin ice
Hamiltonian plus weak transverse terms, indeed constitutes the
predominant part of the low energy effective Hamiltonian, $H_{\rm eff}$,
of TTO. However, as we show below, the microscopic mechanism that leads
to the ``dynamically-induced frustration'' 
and the proposed spin ice assignment for TTO
has heretofore escaped scrutiny. 
We find that frustration and the
spin-ice-like structure of $H_{\rm eff}$ 
dynamically emerge from virtual transitions 
to excited single-ion crystal field (CF) states and, most
importantly, from quantum many-body effects. 
These transitions drastically modify the symmetries of the many-body wave
functions in the low energy sector, leading to a significant
renormalization of the longitudinal (Ising) part of $H_{\rm eff}$. This
renormalization plays a crucial role for materials, such as TTO,
that are naturally-tuned near the boundary between a 
N\'eel ordered phase and the spin ice states. 
In particular, these transitions reposition TTO in 
the spin ice region of coupling parameter space. 
We are led to suggest that TTO 
is a novel quantum 
variant of the classical Ising spin ice materials studied so 
far~\cite{Harris-PRL,Bramwell-JPC,Bramwell-Science,Ramirez,Hertog,Cornelius}.

The main reason for the interest devoted to TTO lies in its failure to 
develop long-range order down to at least 50 mK despite an
antiferromagnetic Curie-Weiss temperature, $\theta_{\rm CW} \sim -14$
K \cite{Gingras-PRB}. 
Similarly to the Dy$_2$Ti$_2$O$_7$ (DTO) and Ho$_2$Ti$_2$O$_7$ (HTO)
spin ices, magnetic Tb$^{3+}$ in TTO possesses a single-ion CF Ising
ground state
doublet with wavefunctions $\vert \Psi_0^\pm\rangle$ where 
$\langle \Psi_0^\pm \vert {\rm J}^z \vert \Psi_0^\pm \rangle$ 
are the only
non-vanishing matrix elements of the ${\bf J}$ angular momentum
operator~\cite{Rosenkranz,Ising}.  
Monte Carlo simulations of a model with
such classical Ising spins~\cite{Hertog} that can only point ``in''
or ``out'' of an elementary tetrahedron~\cite{Ising} and interact via
nearest-neighbor (nn) antiferromagnetic exchange~\cite{Ani-ex,Kao}   
and long-range dipolar couplings  predict,
in dramatic contrast with the 
experimental findings~\cite{Gardner-PRL,TTO-neutron,Gardner-PRB,Gardner-PRB-50mK},
a transition to a four sublattice N\'eel order at
$T_c\sim 1.2~{\rm K}$  \cite{Hertog}.
A key difference between TTO and  spin ices  has so far not
been carefully investigated: 
in spin ices, the excited CF states lie at
an energy~\cite{Gingras-PRB,Rosenkranz} 
several hundred times larger than the exchange and 
dipolar interactions and there is therefore little admixing 
between the CF states induced by the spin interactions. 
This is not necessarily
the case for TTO 
where the first excited doublet lies at only $\Delta\sim 18.7$ K 
above the ground Ising doublet~\cite{Gingras-PRB}. 
It is therefore necessary to 
investigate how the $H_{\rm eff}$ of TTO is 
affected  by virtual quantum mechanical CF excitations.

The Hamiltonian of TTO 
is taken~\cite{Ani-ex} as
$H$ $=$ $H_{\rm cf}+H_{\rm e}+H_{\rm d}$.
$H_{\rm cf}$ is the single-ion CF
Hamiltonian~\cite{Gingras-PRB},
$H_{\rm e}$ $=$ $
{\cal J} \sum_{\langle i,j\rangle} {\rm\bf J}_i \cdot {\rm\bf J}_j$ 
is the nn exchange interaction and 
$H_{\rm d}=
{\cal D}R_{\rm nn}^3 \sum_{(i>j)}
\left[ {{\rm\bf J}_i} \cdot {{\rm\bf J}_j}
 - 3({\rm\bf J}_i \cdot {\hat r}_{ij})
({\rm\bf J}_j \cdot {\hat r}_{ij})
 \right]{|{\bm R}_{ij}|^{-3}} 
$
is the dipole-dipole interaction. 
${\bm R}_{ij}$ 
$\equiv$
${\bm R}_j-{\bm R}_i$
$=$
$\vert {\bm R}_{ij}\vert {\hat  r}_{ij}$,
 where ${\bm R}_i$ is the
position of atom $i$ with total angular momentum ${\rm\bf J}_i$. ${\cal J}$ 
is the nn exchange coupling with the convention here
that ${\cal J}>0$ is antiferromagnetic.
${\cal D}=(\mu_0/4\pi) (g \mu_{\rm B} )^2 /R_{\rm nn}^3$ 
is the dipolar coupling, and $g=3/2$ is the Lande factor for Tb$^{3+}$.
$R_{\rm nn}=3.59 \AA$  is the nn distance,
giving ${\cal D}\approx 0.0315$ K ~\cite{Gingras-PRB}.
To introduce the single-ion wavefunctions which become
admixed by the spin interactions, $H_{\rm int}=H_{\rm e}+H_{\rm d}$, 
we focus on the essential part of $H_{\rm cf}$:  its doublet ground states, 
$\vert \Psi_0^{\pm}\rangle$, and its lowest excited doublet states, 
$\vert \Psi_{\rm e}^\pm \rangle$, at an energy $\Delta = 18.7$ K above 
$\vert \Psi_0^{\pm}\rangle$.
The excited states above $\Delta$  do not lead to qualitatively different 
new physics.
Tb$^{3+}$ has orbital angular momentum L=3, spin S=3, and total angular momentum
${\rm\bf J}={\rm\bf L}+{\rm\bf S}$ with J=6.
We express  
$\vert \Psi_0^{\pm}\rangle$ and $\vert \Psi_{\rm e}^\pm \rangle$  
in terms of the eigenstates 
$\vert  {\rm J} =6, m_{\rm J}\rangle$  
of ${\rm J}_z$ within the fixed ${\rm J}=6$ manifold.
Exact diagonalization of $H_{\rm cf}$ using the 
CF parameters taken from Ref.~[\onlinecite{Rosenkranz}]
for Ho$_2$Ti$_2$O$_7$ but rescaled
 for Tb$_2$Ti$_2$O$_7$
gives:
%
$\vert \Psi_0^\pm \rangle$=
$\alpha_4\vert \pm 4 \rangle 
\pm
\alpha_5\vert \mp 5 \rangle 
+ 
 \alpha_2\vert \mp 2 \rangle
\pm
 \alpha_1\vert \pm 1 \rangle$
and
$\vert \Psi_{\rm e}^\pm \rangle$=
$\pm \beta_5\vert \pm 5 \rangle
+
\beta_4\vert \pm 4 \rangle
+ 
     \beta_2\vert \pm 2 \rangle
\pm
\beta_1\vert \mp 1 \rangle$.
With $\{\alpha_4,\alpha_5,\alpha_2,\alpha_1 \}$=
     $(0.9402,-0.2958,0.1389,0.0960)$ 
and
$\{\beta_5,\beta_4,\beta_2,\beta_1\}$=
$(0.9271,0.3117,-0.1917,0.0809)$.





{\it Exact diagonalization $-$ single tetrahedron.}
Since the spin correlations in TTO never exceed 
a length scale much beyond a single tetrahedron~\cite{Gardner-PRB}, 
we first consider a 
simple model to describe 
TTO
which consist of non-interacting tetrahedra.   
Such an approximation of
independent tetrahedra explains semi-quantitatively several
bulk properties of the classical Heisenberg pyrochlore
antiferromagnet model~\cite{Berlinsky}.
The same  approximation also accounts qualitatively
well for the spin-spin correlations~\cite{Gardner-PRB} 
which, by incorporating transverse spin fluctuations~\cite{Kao,Enjalran-PRB}, 
captures the rough features of the neutron scattering of TTO~\cite{Gardner-PRB}.
Our aim in using this approximation is to expose
the general effects of virtual CF excitations on $H_{\rm eff}$.
We ignore the long-range dependence of the dipole-dipole interactions
in $H_{\rm dip}$ since it is now well understood that it is
the nn contribution of the dipolar 
interactions that predominantly 
controls the transition 
from the
 N\'eel phase to the spin ice state~\cite{Gingras-CJP,Isakov}.
Henceforth, we set  ${\cal D}=0.0315~{\rm K}$  and fix the values of 
$\{\alpha_{m}\}$ and $\{\beta_{m}\}$ to those listed above.
We then treat  ${\cal J}$ and $\Delta$ as independent tunable parameters
in order to expose the essential physics at play 
near the N\'eel $-$ spin ice boundary.


\begin{figure}[t]
\includegraphics[width=9cm]{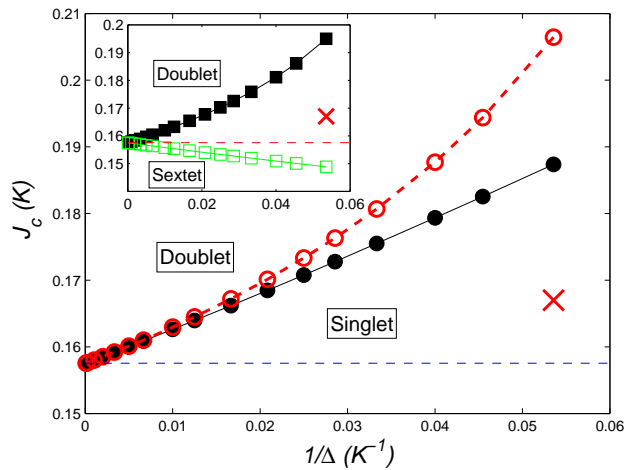}


\caption{
(Color online).
${\cal J}-\Delta$ phase diagram of a single tetrahedron.
TTO has ${\cal J}=0.167$ K and $\Delta=18.7$ K (cross symbol).
Main panel:
The boundary ${\cal J}_c(1/\Delta)$ (filled circles) separates
a  ground state
singlet (${\cal J}<{\cal J}_c$) from a ground state
doublet
(${\cal J}>{\cal J}_c$).
The open circles show the same boundary, but as predicted by
exact diagonalization of $H_{\rm eff}$.
Inset: neglecting transverse terms in $H_{\rm eff}$ ($\lambda=0$),
the filled squares show the renormalized classical sextet-doublet
boundary set by the condition $J_{ij}^{zz}({\cal J},1/\Delta)$ = 0.
The open squares  show the {\it incorrect}
sextet-doublet boundary
predicted when $J_{ij}^{zz}$
for pair $\langle i,j\rangle$
in $H_{\rm eff}$
is obtained
by ignoring contributions
in $PHRHP$ coming from (intermediate) excited states $\vert \Psi_{\rm e}^\pm\rangle$
that belong to the two other Tb$^{3+}$ ions 
($k$ and $l$) on the tetrahedron.
}
\end{figure}


Diagonalizing 
$H_{\rm int}$
for a single tetrahedron within the 
space of the 4$^4$=256 CF states,
we obtain the zero temperature ${\cal J}-\Delta$
phase diagram shown in Fig. 1. 
For the classical
Ising limit (${1/\Delta=0}$), we recover the
transition between ``N\'eel order''
(all-in/all-out, two-fold degenerate, ${\cal J}>5{\cal D}$) 
and a spin
ice manifold 
(two-in/two-out, six fold degenerate, ${\cal J}<5{\cal D}$)
at 
$J_{\rm nn}=D_{\rm nn}$, where 
${J}_{\rm nn}\equiv 
\frac{1}{3} {\cal J}\vert \langle \Psi_0^+
\vert {\rm\bf J}^z \vert\Psi_0^+\rangle\vert^2$
and $D_{\rm nn}\equiv \frac{5}{3}{\cal D}\vert 
\langle \Psi_0^+\vert {\rm\bf J}^z \vert\Psi_0^+\rangle\vert^2$ 
Refs.~[\onlinecite{Hertog,Melko}].
A classical Ising model~\cite{Hertog,Gingras-PRB}
places TTO {\it above} the classical 
$J_{\rm nn}=D_{\rm nn}$ boundary~\cite{Hertog,Melko}
 (horizontal dashed line), i.e. in the N\'eel state~\cite{Gingras-PRB}.
However, for ${1/\Delta>0}$, quantum fluctuations,  due to the
admixing of 
$\vert \Psi_0^\pm \rangle$ with $\vert \Psi_{\rm e}^\pm \rangle$ 
via $H_{\rm int}$,  
become increasingly important, as shown by the renormalized 
${\cal J}_c(1/\Delta)$ boundary in Fig. 1 (filled circles). 
This boundary separates quantum variants of the classical phases and 
TTO, with ${\cal J}=0.167$ K  
and $\Delta=18.7$ K (cross symbol in Fig. 1),
is now {\it deeply} repositioned in the singlet regime, 
i.e. is a {\it quantum spin ice}.
The quantum spin ice state for ${\cal J} < {\cal J}_c$
is a singlet predominently 
built of the fully symmetrized 6 two-in/two-out otherwise
degenerate classical spin ice states whose degeneracy 
is lifted by quantum effects. The ground state
also contains a small 
(of order $1/\Delta$) 
spectral weight contribution from the excited CF states.
The ground singlet
is accompanied by a low energy spectrum of 15 excited states that consists of
three triplets and three doublets spanning an energy band $\delta W\approx$ 0.5 K
above the ground state, and which
is separated by a gap of 16~K from other high energy states.
Above the boundary ${\cal J}_c(1/\Delta)$, 
the ground state is an all-in/all-out doublet 
similarly to the classical $1/\Delta\rightarrow 0$ limit~\cite{Bramwell-JPC,Hertog}.

{\it Single-tetrahedron model $-$ effective Hamiltonian.}
To shed light on the virtual CF excitation
channels leading to the
${\cal J}_c(\Delta)$ above, we construct
an effective $S=\frac{1}{2}$
anisotropic Hamiltonian, $H_{\rm eff}$. Using
second order perturbation theory~\cite{Lindgren} in $1/\Delta$,
we have
$H_{\rm eff} = PHP+PHRHP$,  with
$P  = \sum_{\alpha} \vert \Phi_{0,\alpha}\rangle \langle \Phi_{0,\alpha}\vert$ and
$R = \sum_{\beta} 
{
\vert \Phi_{{\rm e},\beta}\rangle 
({E_0 - E_\beta})^{-1}
\langle \Phi_{{\rm e},\beta}\vert
}
$, 
where $E_0=\langle \Phi_{0,\alpha}\vert H_{\rm cf}\vert  \Phi_{0,\alpha}\rangle$ and
$E_\beta=\langle \Phi_{{\rm e},\beta}\vert H_{\rm cf}\vert  \Phi_{{\rm e},\beta}\rangle$.
Here $\{\vert \Phi_{0,\alpha} \rangle\}$ are 
the $2^4=16$ states constructed as direct products of the
non-interacting single ion $\vert \Psi_0^\pm\rangle$ 
CF doublet ground states of $H_{\rm cf}$.
The $\vert \Phi_{{\rm e},\beta}\rangle$  are
the remaining 4$^4$-16=240 states.
We recast $H_{\rm eff}$
in the form of an effective anisotropic $S=\frac{1}{2}$
spin Hamiltonian in the
individual local [111] spin $\sigma_i^z$ basis~\cite{Ising}:
$H_{\rm eff}
= \sum_{\langle i,j\rangle;\mu,\nu}
 J_{i,j}^{\mu\nu} \sigma_i^\mu \sigma_j^\nu$,
where $\mu,\nu$ are spin component indices, $\sigma_i^\mu$ are Pauli matrices, and
$J_{i,j}^{\mu\nu}$ are the effective anisotropic coupling constants.
A constant energy term has been dropped from $H_{\rm eff}$,
while the one-site $J_i^\mu\sigma_i^\mu$ terms
get eliminated by the symmetry of a tetrahedron.
Figure 1 shows the singlet-doublet boundary predicted by
$H_{\rm eff}$ (open circles).


In order to expose the most important aspects of $H_{\rm eff}$,
we write it as
$H_{\rm eff}=\sum_{<i,j>}
J_{ij}^{zz}\sigma_i^z\sigma_j^z$ $+$ $\lambda\sum_{<i,j>;\mu\nu}
J_{ij}^{\mu\nu}(1-\delta_{\mu z}\delta_{\nu z})
\sigma_i^\mu\sigma_j^\nu $ 
with $J_{ij}^{\mu\nu}=J_{ij}^{\mu\nu}({\cal J},1/\Delta)$
and with the perturbation parameter $\lambda$ ultimately set to $\lambda=1$.
$H_{\rm eff}$ contains transverse 
(non-Ising)
$J_{ij}^{\mu\nu}(1-\delta_{\mu z}\delta_{\nu z})
\sigma_i^\mu\sigma_j^\nu$
terms where, for $\Delta =18.7~{\rm K}$, 
the largest of the transverse $J_{ij}^{\mu\nu}$ is
approximately 50\% of the Ising $J_{ij}^{zz}$ coupling.
To generate these terms via the nonvanishing
matrix elements of ${\rm J}_i^z$ and ${\rm J}_i^\pm$ 
between $\vert \Phi_{0,\alpha} \rangle$ and
$\vert\Phi_{{\rm e},\beta}\rangle$ in $PHRHP$, it
is important to retain more than
the predominant $\alpha_4\vert \pm 4 \rangle$ and
$\beta_5\vert\pm 5\rangle $ components in 
$\vert \Psi_0^\pm\rangle$ 
and $\vert \Psi_{\rm e}^\pm\rangle$.
The term $PHP$ corresponds to the classical [111] Ising model with
nn exchange and dipolar interactions~\cite{Hertog,Gingras-PRB}. 
It is by accident that 
${\cal J}/{\cal D}$ 
has a specific value such that $PHP$
almost vanishes for TTO~\cite{Hertog,Melko}, hence allowing an opportunity 
for the resurgence of quantum
effects via $PHRHP$ in $H_{\rm eff}$.
The contribution
of $PHRHP$ to $J_{ij}^{zz}\sigma_i^z\sigma_j^z$ is
{\it ferromagnetic}, 
and hence competes with the antiferromagnetic classical  
$PHP$ Ising term and 
brings back frustration in TTO.
Neglecting momentarily the quantum transverse terms ($\lambda\rightarrow 0$),
the change of sign of $J_{ij}^{zz}({\cal J},1/\Delta)$
controls the transition from a (spin ice) two-in/two-out sextet to
an all-in/all-out doublet (filled squares in inset of Fig. 1).
It is a key point of this paper that it is 
the renormalization of the Ising sector of the theory,
$J_{ij}^{zz}({\cal J},0)\rightarrow J_{ij}^{zz}({\cal J},1/\Delta)$,
caused by virtual CF excitations,
that largely determines the ${\cal J}_c(1/\Delta)$ boundary 
(filled circles, main panel) and its upward movement with 
respect to the classical
$J_{ij}^{zz}({\cal J},0)=0$ boundary (horizontal dashed line).
Specifically, compare the curve with filled squares in inset of Fig. 1 
with the curve with filled circles in the main panel, and note the
semi-quantitative agreement.






It is 
important to note that the
virtual 
excitation of an intervening (third) ion $k$,
with angular momentum {\bf J}$_k$ and $H=H_{ik}+H_{kj}$
in $PHRHP$, 
plays a crucial role in the renormalization of
the classical Ising sector
$J_{ij}^{zz}\sigma_i^z\sigma_j^z$ for pair $\langle i,j\rangle$ 
in $H_{\rm eff}$. 
Only by including this ``third body'' contribution 
do we get the correct trend for the ``classically renormalized''
 $J_{ij}^{zz}({\cal J},1/\Delta)=0$ boundary.
Failure to do so gives an incorrect boundary
{\it decreasing} with increasing $1/\Delta$
(curve with open squares in inset of Fig. 1).
Hence, it is the quantum many body aspect of the 
full microscopic quantum 
$H=H_{\rm cf}+H_{\rm e}+H_{\rm d}$ of TTO that produces
the interesting physics here, namely the {\it correct}
renormalization
of the Ising part of its $H_{\rm eff}$ from
an unfrustrated system when $1/\Delta=0$ 
to that of a  frustrated
ferromagnetic 
nn spin  ice model~\cite{Bramwell-JPC}.
The aforementioned transverse (quantum) part of $H_{\rm eff}$ ($\lambda\ne 0$)
lifts the degeneracy of the (classical spin ice)
sextet, giving a singlet ground state for an
independent tetrahedron and 15 excited states within
$\delta W \approx 0.5$ K above the ground state.


\begin{figure}[t]
\centering
\includegraphics[width=7cm]{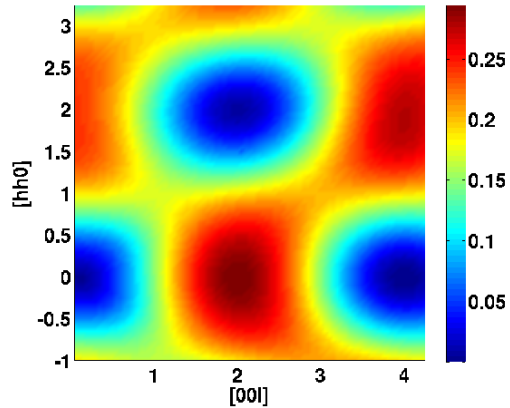}
\caption{(Color online) Theoretical diffuse
neutron scattering intensity 
$I({\bm q})/{\vert F({\bm q})\vert^2}$
for a single 
tetrahedron with four CF states per Tb$^{3+}$ 
in (hhl) plane at 9 K. 
}
\end{figure}

{\it Diffuse neutron scattering.}
The result that $H_{\rm eff}$  for TTO is a nn
{\it ferromagnetic} Ising spin ice model plus transverse terms
might come as a surprise
and be perceived as incompatible with
 neutron scattering measurements~\cite{Gardner-PRB,TTO-neutron,Gardner-PRB-50mK}.
Indeed, the
neutron scattering pattern
of TTO~\cite{Gardner-PRB,TTO-neutron,Gardner-PRB-50mK}  
is qualitatively very different from
that of the HTO ~\cite{Bramwell-PRL-2001} and DTO ~\cite{Fennell}   spin ices.
In TTO,  there is an intensity maximum at 002 in the (hhl) scattering
plane and a second broad maximum at 220.
 In spin ices, there are broad 
maxima at 003 and $\frac{3}{2} \frac{3}{2} 1$ 
~\cite{Bramwell-Science,Bramwell-PRL-2001,Fennell}.
Therefore, the question is whether the above single tetrahedron model
characterized by a ferromagnetic Ising sector in its $H_{\rm eff}$ gives
a diffuse neutron scattering pattern compatible with 
experiment~\cite{Gardner-PRB,TTO-neutron,Gardner-PRB-50mK}.
To address this question we compute the diffuse neutron scattering intensity, $I({\bm q})$,
using standard formulae~\cite{Jensen}:
$I({\bm q})\propto \vert F({\bm q})\vert^2 \sum_{a,b;\alpha,\beta}
[\delta_{\alpha\beta}-q_\alpha q_\beta\vert {\bm q}\vert^{-2}]
S_{\rm diff}^{(a,\alpha;b,\beta)}$
where $a,b$ are the sites on the tetrahedron,
 $\alpha,\beta$ are spin components and
$F({\bm q})$ is the Tb$^{3+}$ form factor.
$S_{\rm diff}^{(a,\alpha;b,\beta)}=\sum_{n,n'}
\langle n | J_{\rm a}^{\alpha} | {n'} \rangle \langle {n'} |J_{\rm b}^{\beta} | {n} \rangle
e^{i{\bm q}\cdot ({\bm r}_b-{\bm r}_a)} e^{-E_n/{k_{\rm B}T} }$
where the states $\{n,n'\}$ 
are those whose energy $E_n$
 falls within the experimental
energy/frequency resolution window of $\sim 4.3$ K  over
which the neutron scattering intensity is energy integrated~\cite{Gardner-PRB}.
These are, incidently,
 the same low energy states that span an energy $\delta W \approx 0.5$ K
above the ground state.
Numerical results for  
$I({\bm q})/\vert F({\bm q})\vert^2$ at $T=9$ K
are shown in Fig. 2.
One finds
a good match in the symmetry of the
theoretical pattern in Fig. 2
with the experimental one
in Fig. 6 of Ref.~\onlinecite{Gardner-PRB}.
These results show that, despite a predominant ferromagnetic nn  Ising component,
$H_{\rm eff}$ possesses sufficient low-energy transverse response (fluctuations)
to account for the symmetry of the diffuse neutron scattering. 
These give, in particular, the intensity maximum at 002 that
 arise from spin fluctuations
transverse to the local [111] Ising directions~\cite{Gardner-PRB,Kao,Enjalran-PRB}.
We propose that a scan of
$I({\bm q})/{\vert F({\bm q})\vert^2}$
along the (hh2) direction may be used to ascertain whether
TTO is indeed in a quantum spin ice state at low temperatures.
Figure 3 shows that 
$I({\bm q})/{\vert F({\bm q})\vert^2}$ along (hh2)
has a broad maximum at h=0 in
the singlet/spin ice regime, 
${\cal J}< 0.187~{\rm K}$ for $1/\Delta=0.053~{\rm K}^{-1}$
(see main panel, Fig. 1),
while it has maxima 
at $h=\pm \delta  ({\cal J})$ in the
doublet
regime for ${\cal J}>0.187~{\rm K}$.
The split hh2 intensity line scan
as a  characterization of the underlying 
(antiferromagnet vs spin ice) ground state 
is sharper the lower the temperature.

Going beyond the
single-tetrahedron approximation, competing anisotropic
interactions further than nearest neighbors  are generated by 
virtual CF excitations. 
These lead to a $H_{\rm eff}$ which, at the classical level displays
various long-range ordered spin ice states depending on ${\cal J}/{\cal D}$. 
Preliminary calculations find a long-range
ordered ${\bm Q}=0$ ferromagnetic (spin ice) state, 
similar to the one recently reported for Tb$_2$Sn$_2$O$_7$~\cite{Mirebeau},
which
competes with the previously identified ${\bm Q}=0$ N\'eel order~\cite{Hertog}
 and the ${\bm Q}=001$ long range ordered spin ice
of the dipolar spin ice model~\cite{Gingras-CJP,Melko}.

\begin{figure}[t]
\centering
\includegraphics[width=4.2cm]{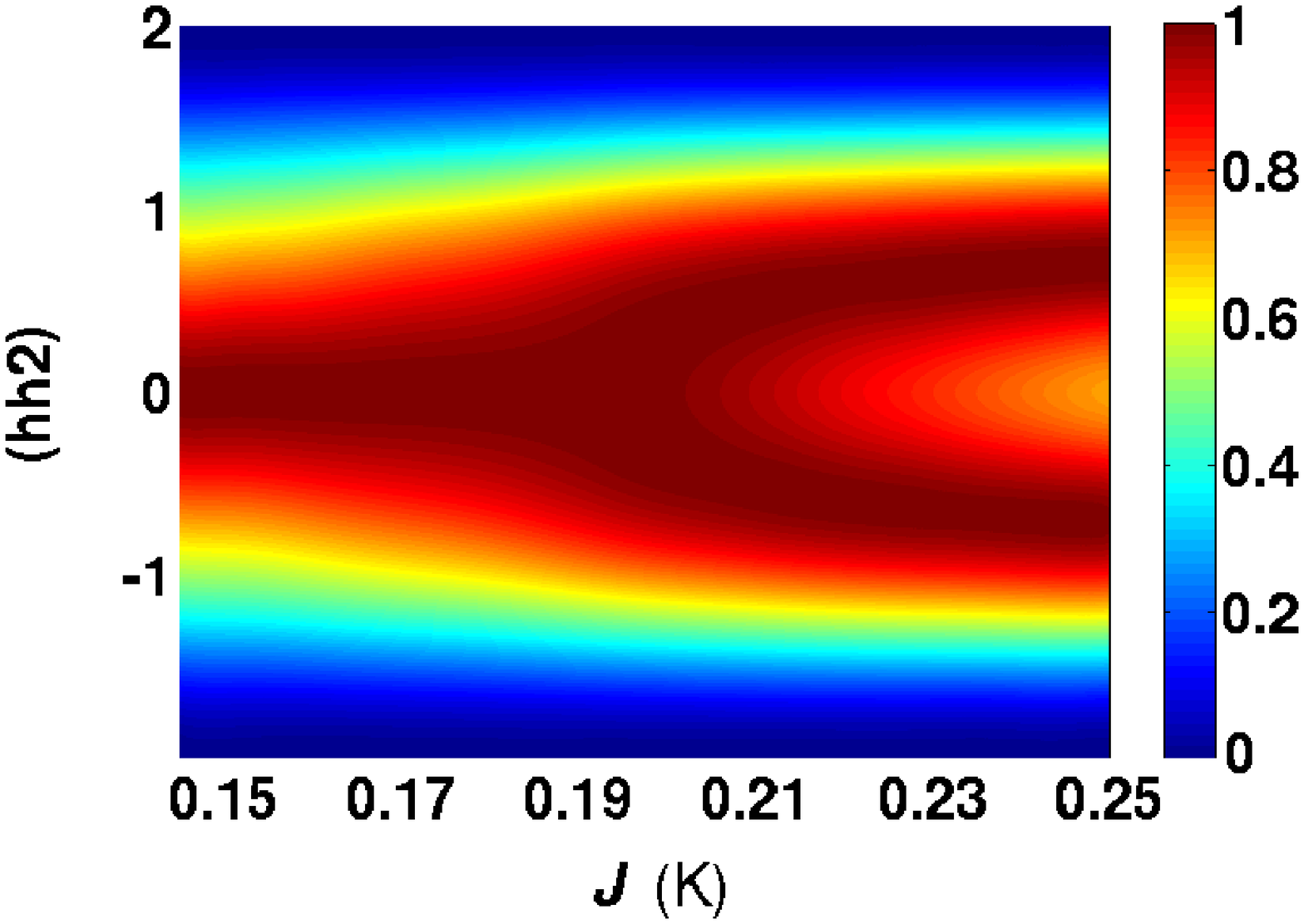}
\includegraphics[width=4.2cm]{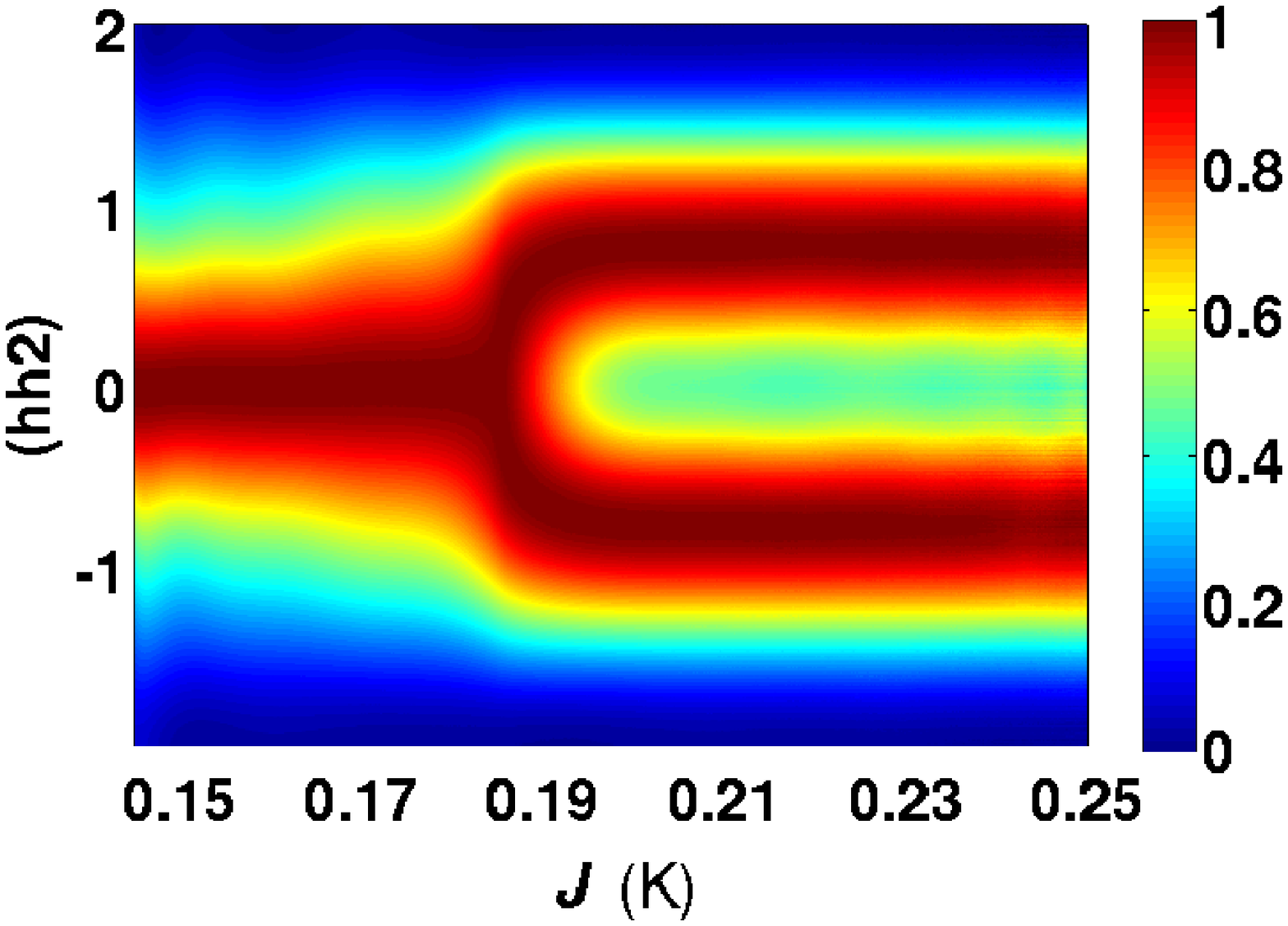}
\caption{(Color online) 
Theoretical diffuse
neutron scattering intensity 
$I({\bm q})/{\vert F({\bm q})\vert^2}$
along ${\bm q}={\rm (hh2)}]$ at 400 mK (left panel)
and 40 mK (right panel)
as a
function of the nearest-neighbor antiferromagnetic exchange
${\cal J}$ for a single
tetrahedron with four CF states per Tb$^{3+}$.
}
\label{hh2scan}
\end{figure}

In conclusion, we have used a simple model of 
non-interacting tetrahedra to describe the low temperature
properties of
of the Tb$_2$Ti$_2$O$_7$ magnetic pyrochlore material.
The present work identifies a new mechanism for 
dynamically-induced frustration in a physical system 
which proceeds via crystal 
field (CF) excitations and quantum many body effects.
More specifically, we uncovered that 
interaction-induced fluctuations among 
otherwise non-interacting single-ion
CF states lead to a renormalization of the low-energy 
effective theory of Tb$_2$Ti$_2$O$_7$ from that of an {\it unfrustrated} [111] pyrochlore
Ising antiferromagnet~\cite{Bramwell-JPC,Hertog}
to a frustrated nearest-neighbor  spin ice model~\cite{Bramwell-JPC,Bramwell-Science}.
The remaining
transverse fluctuations lift the classical ice-like degeneracy and,
at the single-tetrahedron level,
the system is in a quantum mechanically fluctuating spin ice state, or 
{\it resonating spin ice.}
The effects discussed here are likely responsible 
for some of the subtleties underlying the failure of
this material to order at a temperature scale of 1 K~\cite{Hertog,Gingras-PRB,Reviews,Kao}.
Whether the true quantum mechanical ground state of the full lattice model of 
Tb$_2$Ti$_2$O$_7$ is a semi-classical long-range ordered state
with finite quantum spin fluctuations~\cite{Maestro-JPC},
or  a more exotic quantum ground state~\cite{Hermele,Bergman,Pujol}, 
is a challenging but very exciting problem for future studies.

We thank  S. Bramwell, S. Curnoe,
A. Del Maestro, M. Enjalran, T. Fennell, J. Gardner,
B. Gaulin, Y.-J. Kao, S. Rosenkranz, 
J. Ruff and T. Yavors'kii for very useful discussions.
This work was funded by the NSERC
of Canada, the Canada Research Chair Program (Tier I, M.G),
the Province of Ontario
and the Canadian Institute for Advanced Research.
M.G. thanks the U. of Canterbury (UC) for an Erskine Fellowship
and the hospitality of the Dept. of Physics at UC
where part of this work was completed.



\begin{thebibliography}{99}
\bibitem{Ising}
The Ising directions in cubic
pyrochlores are local and correspond to the four cubic $\langle 111 \rangle$
 directions
that intercept elementary tetrahedron unit cells in their middle.
\bibitem{Ani-ex}
A random
phase approximation (RPA) calculations  \cite{Kao}
that consider only nn exchange,
long-range dipolar couplings and the single-ion CF Hamiltonian of Tb$^{3+}$
describe quantitatively well both elastic and inelastic paramagnetic
neutron scattering of TTO \cite{Gardner-PRB}.
Hence, no exceedingly large contribution
from either  
anisotropic electronic exchange,
high order electric multipole-multipole interaction,
exchange interactions beyond nearest neighbors
or virtual phonon exchange 
is  manifestly present in the Hamiltonian $H$ of TTO.
\end{thebibliography}
\end{document}